\documentclass[12pt]{article}
\usepackage{graphicx}
\usepackage{hyperref}
\def\npb#1 #2 #3 #4 {Nucl.~Phys. B {\bf #1}, #2 (#3)#4 }
\def\plb#1 #2 #3 #4 {Phys.~Lett. B {\bf #1}, #2 (#3)#4 }
\def\prd#1 #2 #3 #4 {Phys.~Rev.  D {\bf #1}, #2 (#3)#4 }
\def\prl#1 #2 #3 #4 {Phys.~Rev.~Lett. {\bf #1}, #2 (#3)#4 }

\makeatletter
\def\vereq#1#2{\lower3pt\vbox{\baselineskip1.5pt \lineskip1.5pt
\ialign{$\m@th#1\hfill##\hfil$\crcr#2\crcr\sim\crcr}}}
\makeatother

\newcommand{\newc}{\newcommand}
\newc{\barr}{\begin{eqnarray}}
\newc{\earr}{\end{eqnarray}}
\newc{\beq}{\begin{equation}}
\newc{\eeq}{\end{equation}}
\newc{\ths}{\hskip 0.1cm}

\newcommand{\als}{\alpha_s}   

\newcommand{\gd}{\gamma_\mu}
\newcommand{\gu}{\gamma^\mu}

\def\sss{\scriptscriptstyle}

\def\barp{{\raise.35ex\hbox
{${\sss (}$}}---{\raise.35ex\hbox{${\sss )}$}}}
\def\bdbarp{\hbox{$B_d$\kern-1.4em\raise1.4ex\hbox{\barp}}}
\def\bsbarp{\hbox{$B_s$\kern-1.4em\raise1.4ex\hbox{\barp}}}
\def\barpd{{\raise.35ex\hbox
{${\sss (}$}}--{\raise.35ex\hbox{${\sss )}$}}}
\def\dsbarp{\hbox{$D^{*0}$\kern-1.6em\raise1.5ex\hbox{\barpd}}}
\def\dbarp{\hbox{$D^{0}$\kern-1.1em\raise1.5ex\hbox{\barpd}}} 
\def\kbarp{\hbox{$K^{*0}$\kern-1.6em\raise1.5ex\hbox{\barpd}}}

\begin{document}

\begin{titlepage}
\begin{center}
\today     \hfill    LBNL-50258 \\
~{} \hfill UCB-PTH-02/21  \\
~{} \hfill hep-ph/0205111\\

\vskip .1in

{\large \bf Neutrino mixing and large CP violation in B physics}
\footnote{This work was supported in part by the U.S.
Department of Energy under Contracts DE-AC03-76SF00098, in part by the
National Science Foundation under grant PHY-95-14797.  HM was also
supported by Alfred P. Sloan Foundation.}

\vskip 0.3in

Darwin Chang$^{1,2}$, Antonio Masiero$^{3}$ and Hitoshi Murayama$^{1,4}$

\vskip 0.05in
{\em $^{1}$Theoretical Physics Group\\
     Ernest Orlando Lawrence Berkeley National Laboratory\\
     University of California, Berkeley, California 94720, USA}

\vskip 0.05in
{\em $^{2}$Physics Department and NCTS,\\ 
National Tsing-Hua University, Hsinchu, 30043 Taiwan, ROC}

\vskip 0.05in
{\em $^{3}$Dip. di Fisica "G. Galilei", Univ. di Padova and \\
INFN, Sez. di Padova, Via Marzolo, 8, I-35131 Padua,Italy}

\vskip 0.05in

{\em $^{4}$Department of Physics, University of California\\
     Berkeley, California 94720, USA}

\end{center}

\vskip .1in

\begin{abstract}
We show that in see-saw models of neutrino mass \`a la SUSY $SO(10)$, the 
observed large mixing in atmospheric neutrinos naturally leads to large $b - 
s$ transitions. If the associated new CP phase turns out to be large, this 
SUSY contributions can drastically affect the CP violation in some of the 
$B$ decay channels yielding the $\beta$ and $\gamma$ angles of the
unitarity triangle.  They can even produce sizeable CP asymmetries in some 
decay modes which are not CP violating in the standard model context.  
Hence the observed large neutrino mixing makes observations of low energy 
SUSY effect in some CP violating decay channels potentially promising in 
spite of the agreement between the Standard Model and data in $K$ and $B$ 
physics so far.  
\end{abstract}

\end{titlepage}

\newpage

\section{Introduction}

In the last couple of years we have obtained three major pieces of
information on flavor physics from experiments: large neutrino mixing
in atmospheric neutrinos, existence of ``direct CP violation'' in the
neutral kaon system $\epsilon'/\epsilon$ and observation of CP
violation in $B \rightarrow \psi K_{s}$. Only the first of these three
results clearly calls for new physics beyond the standard model (SM)
for its explanation. As for $\epsilon'/\epsilon$, the theoretical
uncertainties surrounding the SM predictions prevent any firm
conclusion.  Nevertheless, it is reassuring to notice that in the most
familiar and promising extension of the SM, the minimal supersymmetric
SM (MSSM), there is indeed room for a large contribution to
$\epsilon'/\epsilon$ even in the presence of a tiny deviation from
flavor universality in the terms which break SUSY
softly~\cite{hitoshi} . As for the observed CP violation in $B$
physics, the results are in agreement with the SM expectation, but
they leave open the possibility of large CP violating contributions
from new physics (in particular the MSSM) in other decay channels
which should be observable in a near future at $B$ factories or hadron
colliders.

The new physics involved to explain the atmospheric neutrino
observations must provide a mass to neutrinos and guarantee at least
one near-maximal mixing in the leptonic sector.  One of the best candidates
to accomplish these tasks is the see-saw mechanism~\cite{seesaw,seesaw2}. It
has been known for a long time that, at variance with what occurs in
the non-SUSY case, the SUSY version of the see-saw mechanism can
potentially lead to large lepton flavor violating (LFV)
effects~\cite{francesca} \footnote{For recent works on LFV in SUSY
see-saw models, see Ref.~\cite{casas}}. Obviously, then, if one combines
the SUSY see-saw with the idea of some hadron-lepton unification, one may
suspect that the large mixing between second and third generation in
the neutrino case entails not only some large LFV, but also some large
mixing among quarks of the second and third generation which sit
together with the leptons in a GUT multiplet.  This is indeed what may
happen for the right-handed quark supermultiplets in a SUSY $SU(5)$
construction with a see-saw mechanism~\cite{moroi1}

Motivated by the three above experimental observations and the
above-mentioned theoretical considerations in the SUSY context, this
Letter addresses the following question: in a SUSY see-saw context
where neutrino and up-quark couplings are GUT-unified, how large can
the SUSY contributions to CP violation in $B$ physics be? So far it has
been observed that, considering a model-independent parameterization
of the CP violating SUSY contributions in $B$ physics, it is possible,
compatibly with all the existing phenomenological constraints, to
obtain sizeable effects.  However, given that we have now a precise
indication of large LFV effects in the neutrino sector, we find it
timely and important to link this experimental fact to predictions for
$B$ physics in a motivated SUSY GUT context which accounts for the
atmospheric neutrino results.

We find that : 
i) differently from $SU(5)$-like schemes where one has to 
assume the largeness of some neutrino coupling to infer large quark flavor 
violation (FV) 
from the observed large neutrino mixing, in a $SO(10)$-like context the 
link between this latter phenomenon and large $b-s$ transitions is
automatically ensured; 
ii) the mixing $B_s$-$\bar{B}_s$ can receive large 
SUSY contributions comparable, if not larger, than the SM contribution; 
iii) some of the CP violating $B$ decays which yield the $\beta$ and 
$\gamma$ angles of the unitarity triangle are strongly affected by the
presence of the SUSY CP violating contributions, whilst other decays 
which in the SM are predicted to yield the same angles are essentially 
untouched by SUSY; 
iv) there exist some decay channels which do not present any sizeable CP 
asymmetry in the pure SM which develop significant (and hopefully 
observable) CP signals thanks to the large SUSY contributions. The Letter 
shows that, differently from some pessimistic common lore after the CP and 
FCNC results in $K$ and $B$ physics so far, the important experimental finding 
in atmospheric neutrinos yields expectations of sizeable deviations from 
SM in some CP violating $B$ decays in SUSY GUT schemes where neutrino
masses arise from a see-saw mechanism.

\section{Main Point}

The main point of this Letter is very simple.  Consider an $SO(10)$
grand-unified theory, which breaks to $SU(5)$ at, {\it e.g.}\/,
$10^{17}$~GeV.  The Yukawa coupling of ``third-generation'' neutrino
is unified with the large top Yukawa coupling thanks to the $SO(10)$
unification.  However, the large mixing angle in atmospheric neutrino
oscillation suggests that this ``third-generation'' neutrino is
actually a near-maximal mixture of $\nu_\mu$ and $\nu_\tau$.  On the
other hand, the ``third-generation'' charged leptons and down-type
quarks have a relatively large unified bottom and tau Yukawa coupling,
which is diagonal in the $SU(5)$ multiplet that contains $\nu_\tau$ by
definition.  Therefore, the $SU(5)$ multiplet with the large top
Yukawa coupling contains approximately
\begin{equation}
  5^*_3 = 5^*_\tau \cos \theta + 5^*_\mu \sin \theta,
\end{equation}
where $\theta \simeq 45^\circ$ is the atmospheric neutrino mixing
angle, and
\begin{eqnarray}
  5^*_\tau &=& (b^c, b^c, b^c, \nu_\tau, \tau)\\
  5^*_\mu  &=& (s^c, s^c, s^c, \nu_\mu,  \mu).
\end{eqnarray}
It is interesting that even a large mixing in right-handed down
quarks does not appear in CKM matrix simply because there is no
charged-current weak interaction on right-handed quarks.  Therefore
the mixing among right-handed quarks decouples from low-energy
physics. 

However, the mixing among squarks yield observable effects.
The top Yukawa coupling then generates an $O(1)$ radiative correction
to the mass of $\tilde{s}\sin\theta+\tilde{b}\cos\theta$, which leads
to a large mixing between $\tilde{s}$ and $\tilde{b}$ at low energies.
This large mixing in turn generates interesting effects in
$B$-physics.  Examples include: large $B_s$ mixing and CP violation,
different ``$\sin 2\beta$'' in $B_d \rightarrow \phi K_s$ from that in
$B_d \rightarrow J/\psi K_s$ due to CP-violating penguin contribution, and
different ``$\gamma$'' values from different processes.\footnote{In
models of low-energy gauge mediation \cite{GMSB} or anomaly mediation
\cite{AMSB} of
supersymmetry breaking, effects of our interest are simply not present.}

The rest of the paper is devoted to more details of this simple point.
In particular, we demonstrate in the next section that one can write
down semi-concrete models of $SO(10)$ unification which lead to large
$\tilde{b}$-$\tilde{s}$ mixing.  It is important that such models do
not necessarily cause too-large $\mu \rightarrow e\gamma$ or other
dangerous effects.  Then we will discuss detailed consequences of
large $\tilde{b}$-$\tilde{s}$ in $B$ physics.

\section{$SO(10)$}

\subsection{Framework}

We are motivated by $SO(10)$ unification~\cite{georgi}, and we describe
our
assumptions on the unified framework in this section.  When $SO(10)$
is broken to $SU(5)$, we have small mixings among $10$'s responsible
for CKM mixing, while we need large mixings among $\bar{5}$'s to
explain the MNS mixing matrix~\cite{mns} in the neutrino sector.  Then
back in the
$SO(10)$ multiplets, the top
quark comes together with the near-maximal linear combination of
second- and third-generation $\bar{5}$'s.  To the extent that we
ignore all small Yukawa couplings except the top Yukawa coupling, the
only effect of the Yukawa coupling appears in the multiplet
\begin{equation}
  16_{i=3} \ni (t_L, V_{tn} d_{Ln})+ (t_R)^C + U^*_{n3} (d_{Rn})^C + 
(\nu_{L3},
  U^*_{n3} l_{Ln}) + V_{tn} (l_{Rn})^C,
\end{equation}
where we have use indices $i, j, \cdot$ to represent the indices in $Y_u^D$ 
basis while $n, m, \cdot$ are reserved for the basis in which $Y_d$ is 
diagonal.
Here, $V$ is the CKM matrix, and $U$ the MNS matrix (some additional CP 
violating phases will be taken care of later), given the
large $U_{23}$ as evidenced in the atmospheric neutrino data, a
near-maximal linear combination of $s_R$ and $b_R$ experiences the
large top Yukawa coupling in the $SO(10)$ theory.  This simple point
produces a large radiative correction to the soft mass of this linear
combination, which is flavor-off-diagonal in the mass basis of down
quarks.  Therefore we can expect a potentially large effect in $B_s$
mixing, and other related effects in $B$-physics.

Our framework is closely related to that in~\cite{DH} with the
superpotential
\begin{equation}
  W = \frac{1}{2} (Y_u)_{ij} 16_i 16_j 10_u 
  + \frac{1}{2} (Y_d)_{ij} 16_i 16_j 10_d.
\end{equation}
Here, $Y_u$, $Y_d$ are up- and down-type Yukawa matrices, $i,j=1,2,3$
are generation indices, and $10_u$, $10_d$ are Higgs multiplets that
contain $H_u$ and $H_d$ in the MSSM.  

In addition to the above two terms, $W$ must include some further
(renormalizable or non renormalizable) Yukawa coupling responsible for
the right-handed neutrino masses. As usual, this can be achieved
either through $\langle 126\rangle $ or $\langle
\bar{16}\rangle^2/M_{Pl}$.

We need at least two Yukawa matrices $Y_u$ and $Y_d$ to generate 
intergenerational mixings and hence two Higgs
multiplets.  In this sense, this is the ``minimal'' framework of
$SO(10)$ unification. However, this makes both $Y_u$ and $Y_d$ matrices
symmetric.  A symmetric $Y_u$ is acceptable phenomenologically, 
while a symmetric requirement for $Y_d$ turns out to be too strict.
The reason is simply that, in order to accommodate both CKM mixing among 
quarks and MNS mixing among leptons, we need to set
\begin{equation}
  Y_d = \Theta_L V_{CKM}^* Y_d^D \Theta_R U_{MNS} \Theta_\nu,
\end{equation}
in the basis where $Y_u$ and the right-handed neutrino mass matrix are
diagonal (see below). $Y_d^D = \mbox{diag}(Y_d, Y_s, Y_b)$ is the 
positive, diagonalized $Y_d$ matrix.  
$\Theta_L$, $\Theta_R$ and $\Theta_\nu$ are diagonal phase matrices.
Because of this, we will
instead take\footnote{The absence of renormalizable Yukawa coupling to
  $10_d$ could well be a consequence of discrete symmetries.}
\begin{equation}
  W = \frac{1}{2} (Y_u)_{ij} 16_i 16_j 10_u 
  + \frac{1}{2} (Y_d)_{ij} 16_i 16_j \frac{\langle 45 \rangle}{M_{Pl}}
  10_d.
\end{equation}
Because of the combination of the Higgs multiplet $45$, whose VEV $\langle
45 
\rangle \neq 0$ breaks SO(10), and the Higgs in 10, the effective Yukawa 
coupling being either in 10 (symmetric between two
16's) or 120 (anti-symmetric between two 16's) representations, the
matrix $Y_d$ can now have a mixed symmetry.  We imagine that $SO(10)$ is
broken to $SU(5)$ around $10^{17}$~GeV, and this operator is large
enough for the down-type Yukawa matrix.  
Note that we define $V$ and $U$ matrices to be in the CKM form with only one 
CP violating phase each.  
Phases in $\Theta_L$ and $\Theta_R$ are relevant only when the superheavy 
color triplet components of the Higgs multiplet is involved.
$\Theta_\nu$ is relevant for the CP violation in the neutrino sector when 
the Majorana character of the neutrino mass is involved.

Now we break $SO(10)$ to $SU(5)$.  Given a strong hierarchy among
up-type Yukawa couplings and the large top Yukawa coupling, it is
natural to stick to the basis where $Y_u$ is diagonal, $(Y_u)_{ij} =
(Y_u^D)_i \delta_{ij}$.  Further decomposing multiplets under $SU(5)$
as $16_i = 10_i + \bar{5}_i + 1_i$, and keeping only the Higgs multiplets 
$5_u \in 10_u$ and $\bar{5}_d \in 10_d$, we find
\footnote{Right-handed neutrino
  masses arise from the coupling to $SO(10)$-breaking Higgses either
  $\langle 126\rangle $ or $\langle \bar{16}\rangle^2/M_{Pl}$, as
  usual.}
\begin{equation}
  W = \frac{1}{2} (Y_u^D)_{i} 10_i 10_i 5_u 
  + (Y_u^D)_{i} \bar{5}_i 1_i 5_u
  + (Y_d)_{ij} 10_i \bar{5}_j \bar{5}_d
  + \frac{1}{2} M_{ij} 1_i 1_j.
  \label{eq:SU(5)}
\end{equation}

It is clear that, in the absence of the right-handed neutrino mass matrix,
we can eliminate $U_{MNS}$ entirely by changing the basis of
$\bar{5}_i$ in the $SU(5)$ superpotential.\footnote{This, of course,
  is not necessarily true with the soft terms, which is the whole
  point of this paper.}  This is an immediate way to see that the only
effect of $U_{MNS}$ is related to the neutrino mass.

Further breaking $SU(5)$ down to the standard model, the prediction of this
framework is that the Yukawa couplings in the MSSM+N (the MSSM
together with right-handed neutrinos) are
\begin{eqnarray}
  \lefteqn{
    W = (Y_u^D)_{i} Q_i U_i H_u  + (Y_u^D)_{i} L_i N_i H_u 
  }
  \nonumber \\
  & & \hspace{-0.5cm}
+ (V^* Y_d^D \Theta_R U \Theta_\nu)_{ij} Q_i D_j H_d
+ (V^* Y_d^D \Theta_R U \Theta_\nu)_{ij} E_i L_j H_d
  + \frac{1}{2} M_{ij} N_i N_j,
  \label{eq:MSSM+N}
\end{eqnarray}
where we had absorbed the phases in $\Theta_L$ into $Q_i$ and $E_i$.
$Y_d$ is diagonalized by a bi-unitary rotation where the matrix acting on
the left side represents the relative rotation of the left-handed down
quarks with respect to the left-handed up quarks in the basis where the up
quark mass matrix is diagonal. Hence, such matrix is just the usual CKM
mixing matrix. The matrix acting on the right side represents the rotation
to be performed on the left-handed leptons to go the mass basis of the
charged leptons.  
It is easy to see that the phase matrix $\Theta_\nu$ can be absorbed into 
the Majorana mass matrix $M$ after redefining $D_j$ and $L_j$, and, the 
phase matrix $\Theta_R$ can be absorbed into $(UD)$ multiplet or $(EV^*)$ 
multiplet. The phase matrices $\Theta_{L,R}$ are irrelevant 
as long as the colored triplet Higgs boson can be ignored as emphasized 
before.  
Hence such matrix $U$ is to be identified with the
neutrino mixing matrix  $U_{MNS}$ if we are in a basis where the physical
light neutrinos are mass eigenstates.  For this to happen, given that the
neutrino Yukawa coupling matrix $Y_u$ is diagonal, we have to assume that
simultaneously also the right-handed neutrino mass matrix $M$ is diagonal.
Hence throughout our discussion we are taking  {\it $Y_u$ and $M$
simultaneously diagonal}\/. Such a situation could result from simple
$U(1)$ family symmetries. As we will comment below, in a scheme a la
$SO(10)$ with hierarchical
$Y_u^D$ and right-handed neutrino masses, the choice of having such
simultaneous diagonalization looks rather plausible.\footnote{For a
discussion of neutrino masses and mixings based on the simplest $SO(10)$
mass relations and the see-saw mechanism, see the work of
Ref.\cite{wyler}}

The similarity of the charged lepton and down-quark Yukawa matrices is
well-known phenomenologically.  Quantitatively, the relation
$m_b=m_\tau$ could be indeed true at the unification scale, while $m_s
= m_\mu$, $m_e = m_d$ are a factor of about three off.  Here we take the
point of view that the factors of three can be remedied by small
$SU(5)$-breaking effects of the framework and do not worry about it.
Clearly, lower-generation Yukawa couplings are subject to more
corrections simply because their sizes are small.  The $B$-physics
signatures we will discuss do not depend on such details as we end up
ignoring all Yukawa couplings except that of the top quark.  It is
important to notice that the order of left- and right-handed fields is
the opposite between $Q_i D_j$ and $E_i L_j$ couplings.

The important outcome of this framework is the (approximate) equality of the
neutrino and up-quark Yukawa matrices.  The light neutrino masses,
after integrating out the right-handed neutrinos in
Eq.~(\ref{eq:MSSM+N}), are given by the superpotential
\begin{equation}
  W = \frac{1}{2} (Y_u^D)_{i} (M^{-1})_{ij} (Y_u^D)_{j} (L_i H_u) (L_j H_u).
\end{equation}
Since we assume that $M$ is also diagonal in the same basis, this leads to 
the light Majorana neutrino mass matrix 
$(m_\nu)_{nm} = (Y_{ui}^{D 2}/2M_i) U_{ni}^* e^{- i \delta_i} U_{mi}^*$, 
where $e^{i \delta_i}$ is the phase of $M_i$. 
The immediate conclusion is that the right-handed neutrino Yukawa
matrix must be roughly doubly hierarchical compared to the up-quark
Yukawa matrix.  Phenomenologically, the Large Angle MSW solution is
the most promising solution to the solar neutrino problem.  Then the
two mass splittings~\cite{SNO}
\begin{eqnarray}
  \Delta m_\oplus^2 &\simeq& 3 \times 10^{-3}~\mbox{eV}^2, \\
  \Delta m_\odot^2 &\simeq& 0.3\mbox{--}2 \times 10^{-4}~\mbox{eV}^2,
\end{eqnarray}
are not very different, especially after taking their square root.  On
the other hand, the up-quark Yukawa matrix has a strong hierarchy $Y_u
\ll Y_c \ll Y_t$.  To obtain similar mass eigenvalues between the
largest and the 2nd largest eigenvalues as suggested by data, we need
$Y_c^2 / M_2 \sim 0.2 Y_t^2 / M_3$.  Moreover, the basis where the
$Y_u$ matrix is diagonal must be strongly correlated to the basis
where the right-handed neutrino masses are diagonal to achieve this.
The simplest possibility is to assume their simultaneous
diagonalization, as we said above.  Note also that all three physical
CP violating phases associated with the light $3\times 3$ Majorana
mass matrix are present in this model as free parameters.

At GUT scale, the top quark mass and the largest neutrino Dirac mass
are equal.  As a result the heaviest neutrino mass is $m_t^2/M_3$.
From the recent fit to the Super-Kamiokande neutrino data and assuming
non-degenerate neutrino masses, one has $m_{\nu_3} \sim 0.05$~eV.  For
$m_t \sim 178$~GeV, this corresponds to $M_3$ of roughly $10^{15}$~GeV
slightly below the GUT scale as expected.  It is very interesting to
see that the $SO(10)$ model ties up neutrino mass, top mass and GUT
scale nicely.


\subsection{Effects on Soft Masses}

The size of the radiative corrections on the SUSY soft masses induced by the
neutrino Yukawa couplings and their possible consequence on low-energy
flavor physics had been studied within the $SU(5)$ unification in
Refs.~\cite{moroi1,moroi2,lepton,susyb}.  Following these papers, we shall
assume that above some GUT unification scale, the SUSY breaking
parameters are universal and can be parameterized by the universal
scalar mass $m_0$, the universal $A$-parameter $a_0$ which is the
ratio of the SUSY breaking trilinear scalar interaction to the
corresponding Yukawa couplings, the $B$-parameter entering the scalar bilinear 
term mixing the two Higgs doublets and the universal
gaugino mass $m_G$.
The scale, $M_*$, where these universal SUSY breaking values should be
applied depends on the details of the SUSY breaking mechanism.  Here
we shall simply assume it to be near the Planck scale .

In the context of SUSY $SU(5)$~\cite{moroi1, moroi2}, it was shown 
that, if the right-handed neutrino singlet is introduced to account for the 
data on neutrino oscillation,  large neutrino Yukawa couplings involved in 
 the neutrino Dirac masses can induce large off-diagonal 
mixings in the right-handed down squark mass matrix through renormalization 
group evolution between $M_*$ and $M_{GUT}$.  In addition, the contributions
 to the
scalar masses induced in the running by the   neutrino 
Yukawa couplings will generally be complex with new CP violating phases
unrelated to 
the Kobayashi-Maskawa(KM) phase in Standard Model. The abovementioned mixings 
 can be parameterized as 
$\delta^{R}_{ij} = (m^2_{\tilde{d}_R})_{ij}/ m_{\tilde{q}^2}$ where 
$m_{\tilde{q}^2}$ is the average right-handed down squark mass.
In particular, in Ref.~\cite{moroi1}, it was shown that  the induced 
$\delta^{R}_{12}$ is large enough to account for many of the observed 
CP violating phenomena in the kaon system providing an alternative to the CKM 
interpretation of these data.  In Ref.~\cite{moroi2}, it was shown that 
$\delta^{R}_{13}$ can give rise to a 
CP asymmetry in  
$B_d \rightarrow \phi K_s$
much larger than the KM prediction.  

Here we wish to point out first that the off-diagonal mixing parameter
$\delta_{23}$ is further enhanced in the context of a SUSY $SO(10)$
model.  In the next section we will elaborate on the phenomenological
consequence of large $\delta^{R}_{23}$.  In our case the constraint coming 
from the upper bound on $BR(\mu \rightarrow e\gamma)$
turns out to be less severe than in the $SU(5)$ context.

Due to the larger matter content of the  $SO(10)$ GUT model, the
renormalization group evolution from $M_*$ down to $SO(10)$ breaking
scale, $M_{10}$, is faster than that of the $SU(5)$ model.  The induced
off-diagonal elements in the  SUSY breaking mass matrix of the right-handed down
squarks  $\tilde{d}_R$ are given by (in the basis in which $Y_D$ is
diagonal) 
\beq
[m^2_{\tilde{d}_R}]_{nm} \simeq
-\frac{1}{8\pi^2} \left[ Y^{u\dagger} Y^u \right]_{nm} (3m_0^2+a_0^2)
\left(5 \log \frac{M_*}{M_{10}} + \log \frac{M_{10}}{M_{5}} \right)
\label{eq:msquare}  
\eeq
where $M_5$ is the $SU(5)$ breaking scale and
\begin{equation}
\left[ Y^{u\dagger} Y^u \right]_{nm} 
= \left[ (\Theta_R U Y_u^{D 2} U^\dagger \Theta_R^* \right]_{nm} 
= e^{-i(\phi^{(L)}_m-\phi^{(L)}_n)}
y_t^2 \left[U\right]_{m3}^* \left[U\right]_{n3},
\end{equation}
where $e^{i \phi^{(L)}_n}$ is the phase from $(\Theta_R)_{nn}$.  Note that 
these 
phases are not relevant to any other low energy physics.

To account for the large atmospheric neutrino mixing, the second and
third  entries of the
third row of $U_{MNS}$ should be of order $\sqrt{2}$, while the first
entry,
$U_{3e}$, is severely limited by the CHOOZ experiment:  $|U_{e3}|   
\leq 0.11$.
Hence we obtain:
\begin{equation}
[Y^{u\dagger}Y^u]_{23} = 0.5
e^{-i(\phi^{(L)}_2-\phi^{(L)}_3)}(m_{tG}/178\mbox{ GeV}) 
^2,
\label{eq:delta}
\eeq
where $m_{tG}$ is the top quark mass at $M_{G}$.

 The factor 5 in the 
RG coefficient above the $SO(10)$ breaking scale is due to the
contribution of the loop diagram with $(10, \bar{5}_u)$ multiplets of
$SU(5)$ in the loop which is not present in  $SU(5)$.
They contribute four times more than the usual $(1, 5_u)$
contribution in $SU(5)$, and $\delta_{23}$ can easily be $O(1)$.

Note that $m_{tG}$ can be quite different from the pole mass of about
$m_t \sim 178$~GeV.  The evolution of $m_{tG}$ between $M_*$ and
$M_{G}$ has been discussed in the literature \cite{susygut}.

In $SU(5)$ models, people had assumed that the right-handed neutrino
mass matrix is given by an identity matrix to simplify the analysis.
Given only a small hierarchy between $\Delta m^2_\oplus$ and $\Delta
m^2_\odot$ for Large Mixing Angle MSW Solution, the second-generation
neutrino Yukawa coupling is sizable, with a large mixing with the
first-generation states.  This led to quite stringent constraints from
the processes such as $\mu \rightarrow e\gamma$.  In our $SO(10)$
framework, however, the neutrino Yukawa matrix is as hierarchical as
the up-quark sector, and only the third-generation Yukawa coupling is
significant.  In the third-generation multiplet, the electron state
appears with a suppressed coefficient $U_{e3}$.  Therefore, unlike the
frameworks studied in the literature, contributions to  processes that
involve the first generation (like $\mu \rightarrow e+\gamma$) are
suppressed by this unknown element in
the MNS matrix. In view of this fact, we prefer to focus on flavor
violating processes involving the 
 mixing between second- and third-generation in this paper.  

\section{Consequences in $B$ Physics}

In this Section we present some implications of a large and complex
$\delta^{R}_{23}$ in $B$ physics.  The discussion is semi-quantitiative.
Fully quantitative evaluation of effects in $B$-physics and their
corelations will be discussed elsewhere.\footnote{We stick to mass insertion
formalism for illustrative purposes, but fully quantitative
discussions would require calculations in the mass eigenbasis.}

The diagrammatic contributions of
$\delta^{R}_{23}$ to various
$\Delta B = - \Delta S = 2$
and $\Delta B = - \Delta S = 1$ processes were worked out in detail in
Ref.~\cite{gabbiani}\footnote{For an updated analysis of the
gluino-mediated  SUSY contributions to  the  $B_d$-$\bar{B}_d$ mixing and
to the CP asymmetry in the decay $B \rightarrow J/\psi K_s$ 
 including the NLO
QCD corrections and $B$ coefficients as computed in the lattice instead of
using the vacuum insertion approximation, see Ref.~\cite{ciuchini}}. In
particular a
complex $\delta^{R}_{23}$ can play a major role in CP violating $B$ decays
\cite{ciuchini2,moroi2,lunghi,susyb,dc}. 

\begin{figure}[tbp]
  \centering
  \includegraphics[width=0.45\textwidth]{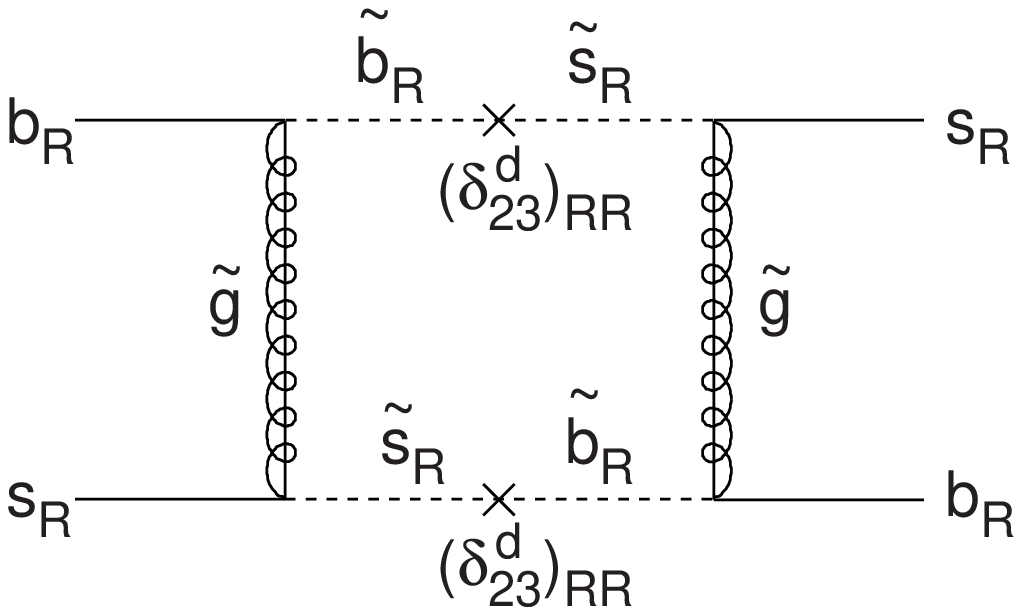}
  \includegraphics[width=0.45\textwidth]{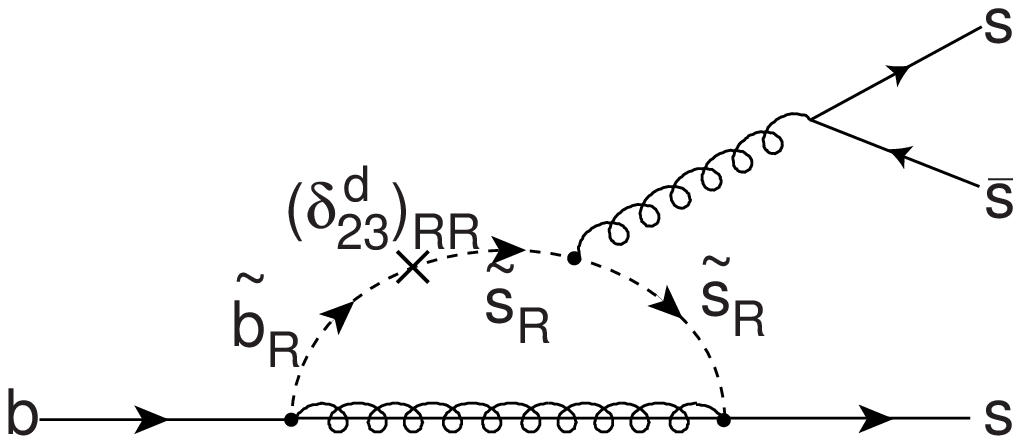}
  \caption{Possible important contributions to $B$-physics from large
    $\tilde{b}_R$-$\tilde{s}_R$ mixing, such as $B_s$-$\bar{B}_s$
    mixing, and SUSY penguin contribution to $B_d \rightarrow \phi
    K_s$ transition.}
\end{figure}

The first effect of a conspicuous $\delta^{R}_{23}$ would be a
large contribution to the 
$\Delta B = - \Delta S = 2$, $B_s$-$\bar{B}_s$ mixing through the operator
$Q_1 = \bar{s}^{\alpha}_R \gd b^{\alpha}_R \bar{s}^{\beta}_{R} \gu 
s^{\beta}_R$
with complex coefficient
\beq
{\cal{H}}_{\it eff}=-\frac{\als^2}{216 m_{\tilde{q}}^2}  
        \left(\delta^d_{23}\right)^2_{RR}  
        \left(  24\,Q_1\,x\,f_6(x) + 66\,Q_1\,\tilde{f}_6(x) \right)
\eeq
where the functions $f_6$ and $\tilde{f}_6$ are defined as in
Ref.~\cite{gabbiani}.

From Eqs.~(\ref{eq:msquare},\ref{eq:delta}) we see that
$\delta^{R}_{23}$ can easily 
be as large as $0.5$, yielding a SUSY contribution to $\Delta M_s$
comparable to that of the SM. Hence, in our scheme  
the operator gives rise to a large $B_s$-$\bar{B}_s$ mixing with a complex
phase 
which is almost unconstrained so far.  In contrast, the SM contribution to 
$B_s$-$\bar{B}_s$ mixing has very small phase (in the usual Wolfenstein 
convention).  This gives rise to many phenomenological consequences as will 
be sampled below.  

\setcounter{footnote}{0}
Secondly, $\delta^{R}_{23}$ gives rise to new contribution to direct 
$B$ decays.
Two categories of contributions are more important.  
The first one is a $\Delta B =1$ box diagram contribution.  The second is 
of the type of electromagnetic or gluonic Penguin contributions.\footnote{Fully
  quantitative analysis would require a careful study of possible
  cancellation between penguin and box diagrams \cite{GKN}.}  There are 
also contribution of electroweak Penguin type which we shall ignore because 
they are generally smaller.

In $B$ decay processes in which the dominant contribution in SM is at the
tree 
level, the additional contribution due to $\delta^{R}_{23}$ can at most be 
a small percentage.  This is true even if the SM contribution comes with 
strong mixing angle suppression \cite{dc} such as $B^\pm \rightarrow D K^\pm$ 
or $\bar{D} K^\pm$. 
However, if the initial state mixing (such as $B_S$ mixing) 
plays a strong role in the phenomena, then $\delta^{R}_{23}$ can significantly 
alter the phenomenology.

In $B$ decay processes in which the dominant contributions involve one
loop 
contributions, such as the Penguin diagrams, one should expect large 
additional contribution due to the $\delta^{R}_{23}$ in both amplitude and CP 
asymmetry.

As an application of the above analysis, we can roughly classify the 
phenomonology 
into three categories.
\renewcommand{\theenumi}{(\alph{enumi})}
\begin{enumerate}
\item Measurements of the $\beta$ angle of the unitarity triangle.
  The leading mode, $B \rightarrow J/\psi K_s$, has large phase from
  the initial state $B_d$ mixing (in Wolfenstein convention), and
  large real tree-level decay amplitude.  It therefore does not receive
  significant contribution from $\delta^{R}_{23}$.  However, other
  modes, such as $B_d \rightarrow \phi K_s$ ($\bar{b} \rightarrow
  \bar{s} s \bar{s}$), which, in SM, measures the same $\beta$, can
  now receive large additional contribution from $\delta^{R}_{23}$
  through the box and Penguin diagrams \cite{moroi2,lunghi}.  The fractional
  phase difference $r_\beta = (\beta(J K_s) - \beta(\phi K_S))/\beta(J
  K_s)$ is a measure of $\delta^{R}_{23}$ that can be as large as
  $50 \%$.
\item Measurements of the $\gamma$ angle of the unitarity triangle.
  One class of popular measurements on $\gamma$ involves $B_s$
  decays \cite{review}.  In the SM, the $B_s$ mixing is real to a very
  good
  approximation.  Therefore any measurements of $\gamma$ using $B_s$
  decays will be strongly affected by $\delta^{R}_{23}$.  For example,
  in $\bar{B}_s \rightarrow (D_s)^- K^+$ the CP asymmetry is due to
  the interference of the $B_s$ decays, which has real amplitude in
  SM, and the $\bar{B}_s$ decay with complex amplitude after the
  $B_s$-$\bar{B}_s$ mixing.  The two decays are roughly of equal
  magnitude and the phase of $\bar{B}_s$ is exactly $\gamma$ in SM.
  With $\delta^{R}_{23}$, even the $\bar{B}_s \rightarrow B_s
  \rightarrow (D_s)^- K^+$ develops a large phase due to additional
  complex contribution to the mixing.  The phase is proportional to
  $\mbox{Arg}(M_{12}(\delta^{R}_{23})/(M_{12}(\delta^{R}_{23}) +
  M_{12}({\rm SM})))$ where $M_{12}({\rm SM})$ and
  $M_12(\delta^{R}_{23})$ are the $B_s$ mixing amplitude of the SM and
  that due to $\delta^{R}_{23}$ respectively \cite{dc}.
  
  Another class of measurement on $\gamma$ are using charge $B$
  decays.  For example in $B^+ \rightarrow D^0 K^+$ or $\bar{D}^0 K^+$,
  $\gamma$ is measured through the interference between the two quark
  level processes $b\to c \bar{u}s$ and $b\to u \bar{c} s$. The
  decay chains of $c$ or $\bar{c}$ in the final state produce a $D^0$
  or $\overline{D}^0$ mesons respectively. The two contributions
  interfere if both $D^0$ or $\overline{D}^0$ decay to the same final
  state $f_D$ and have a relative phase $\gamma$.  Here, $f_D$ is one
  of the states that both $D$ and $\bar{D}$ can decay into, such as
  $K^- \pi^+$ or CP eigenstates $K^+ K^-$, $\pi^+ \pi^-$, $K_s \pi^0$
  or $K_s \phi$ \cite{dc}.  In this measurement, the role played by
  $\delta^{R}_{23}$ is negligible, so it measures the same value as in
  SM.  Therefore, just like $\beta$ measurements, by comparing
  $\gamma$ measurements in $B_s$ and in $B^\pm$ decays, one can get a
  measure of $\delta^{R}_{23}$.  In principle, by comparing $r_\beta$
  with $r_\gamma$, which is similarly defined, one can get a strong
  evidence of the existence of large $\delta^{R}_{23}$.
  
\item Decays which are expected to be essentially CP conserving in the
  SM.  Some of decays may have large CP asymmetry due to the existence
  of $\delta^{R}_{23}$.  Examples: $B_s \rightarrow J \phi$ or $B_s
  \rightarrow (D_s)^+(D_s)^-$ or $B \rightarrow X_s \gamma$.
\end{enumerate}

\section{Conclusion}

In this Letter, we pointed out that a large mixing between $\nu_\tau$
and $\nu_\mu$ as observed in atmospheric neutrino oscillation may lead
to a large mixing between $\tilde{b}_R$ and $\tilde{s}_R$ because they
belong to the same $SU(5)$ multiplets.  This occurs naturally in
$SO(10)$ grand unified models which we have described in detail.  These
models do not give rise to dangerously large $\mu \rightarrow e\gamma$
and similar processes which involve 
the first generation, given the current limit on $U_{e3}$ from
reactor-neutrino experiments.   A large mixing between $\tilde{b}_R$
and $\tilde{s}_R$ leads to interesting effects in $B$-physics, such
as large and CP-violating $B_s$ mixing, different ``$\sin
2\beta$'' between $B_d \rightarrow \phi K_s$ and $J/ \psi K_s$,
different ``$\gamma$'' from various measurements, and CP asymmetry in
$B_s \rightarrow J\phi$, $(D_s)^+ (D_s)^-$.

{\it Acknowledgements}.

HM was supported in part by the Director, Office of
Science, Office of High Energy and Nuclear Physics, Division of High
Energy Physics of the U.S. Department of Energy under Contract  
DE-AC03-76SF00098 and in part by the National Science Foundation
under grant PHY-0098840.  DC is supported by a grant from NSC
of ROC and LBL. DC 
wishes to thank the Theory Group at LBL and the Theory Group at SLAC for 
hospitality while this work is in progress and Lincoln Wolfenstein for 
discussion. AM thanks the UCB and LBL Theory Groups for kind
hospitality and support at the early stages of this work. 

\section*{Note Added}

After this manuscript is submitted to the arXiv, we learned from Jogesh
Pati that a similar idea by K.S. Babu and J. Pati was mentioned in his
talk at the CasCais, Portugal School (July 2000), the abstract of
which appeared in the Proceedings edited by Branco, Shafi and
Silva-Marcos, Pages 215. 

\bibliographystyle{unsrt}

\end{document}